\documentclass[aps,prb,twocolumn,superscriptaddress,showpacs]{revtex4-2}
\usepackage{amsmath,amssymb,wasysym,graphicx}
\pdfoutput=1
\usepackage[utf8]{inputenc}
\usepackage[T1]{fontenc}
\usepackage{xcolor}

\IfFileExists{newtxtext.sty}
   {\usepackage{newtxtext,newtxmath}}
   {\IfFileExists{stix.sty}
      {\usepackage{stix}}
      {\IfFileExists{mathptmx.sty}
      {\usepackage{mathptmx}}{} } }

\usepackage{textcomp}

\usepackage{bm}

\IfFileExists{siunitx.sty}{\usepackage{booktabs,siunitx}}{}

\usepackage{color}
\definecolor{LinkColor}{rgb}{0.256,0.439,0.588}
\usepackage{hyperref}
\hypersetup{
   colorlinks=true,
   citecolor=LinkColor,
   linkcolor=LinkColor,
   urlcolor=LinkColor
}

\usepackage{pifont}
\usepackage[normalem]{ulem}

\usepackage{multirow}

\begin{document}

\title{Dirac fermions with plaquette interactions. I. SU(2) phase diagram with Gross-Neveu and
deconfined quantum criticalities}

\author{Yuan Da Liao}
%\email{ydliao@fudan.edu.cn}
\affiliation{State Key Laboratory of Surface Physics, Fudan University, Shanghai 200438, China}
\affiliation{Center for Field Theory and Particle Physics, Department of Physics, Fudan University, Shanghai 200433, China}

\author{Xiao Yan Xu}
%\email{xiaoyanxu@sjtu.edu.cn}
\affiliation{Key Laboratory of Artificial Structures and Quantum Control (Ministry of Education), School of Physics and Astronomy, Shanghai Jiao Tong University, Shanghai 200240, China}

\author{Zi Yang Meng}
\email{zymeng@hku.cn}
\affiliation{Department of Physics and HKU-UCAS Joint Institute of Theoretical and Computational Physics, The University of Hong Kong, Pokfulam Road, Hong Kong SAR, China}

\author{Yang Qi}
\email{qiyang@fudan.edu.cn}
\affiliation{State Key Laboratory of Surface Physics, Fudan University, Shanghai 200438, China}
\affiliation{Center for Field Theory and Particle Physics, Department of Physics, Fudan University, Shanghai 200433, China}
\affiliation{Collaborative Innovation Center of Advanced Microstructures, Nanjing 210093, China}
\date{\today}

\begin{abstract}
We investigate the ground state phase diagram of an extended Hubbard model with $\pi$-flux hopping term at half-filling on a square lattice, with unbiased large-scale auxiliary-field quantum Monte Carlo simulations.
As a function of interaction strength, there emerges an intermediate phase which realizes two interaction-driven quantum critical points,
with the first between the Dirac semimetal and an insulating phase of weak valence bond solid (VBS) order,
and the second separating the VBS order and an antiferromagnetic insulating phase.
These intriguing quantum critical points are respectively bestowed with Gross-Neveu and deconfined quantum criticalities, and the critical exponents $\eta_\text{VBS}=0.6(1)$ and $\eta_\text{AF}=0.58(3)$ at deconfined quantum critical point satisfy the CFT Bootstrap bound.
We also investigate the dynamical properties of the spin excitation and find the spin gap open near the first transition and close at the second. The relevance of our findings in realizing deconfined quantum criticality in fermion systems and the implication to lattice models with further extended interactions such as those in quantum Moir\'e systems, are discussed.
\end{abstract}

\maketitle

{\it Introduction.---}
The Landau-Ginzburg-Wilson (LGW) paradigm of phases and their transitions is one of the cornerstone of modern condensed matter physics~\cite{Statistical1980,wilsonRenormalization1974}, in which, the phase transition could be understood in terms of symmetry breaking and the establishment of order parameters.
According to LGW, the transition between two ordered phases with spontaneously broken symmetries should either be first-order or through an intermediate phase.
However, new transitions between novel quantum states that are beyond the LGW are accumulating in recent years.
For example, Senthil {\it et al.} proposed that a continuous quantum phase transition between the antiferromagnetic (AF) order and the valence bond solid (VBS) order could exist~\cite{senthilDeconfined2004,senthilQuantum2004}, dubbed as the deconfined quantum critical point (DQCP).
Strong evidence of DQCP in spin-1/2 model on square lattice has been first shown in the J-Q model by Sandvik~\cite{sandvikEvidence2007,louAntiferromagnetic2009}.
Subsequently, other numerical examples and new theoretical understanding of DQCP have been developed in quantum spin models~\cite{blockFate2013,nahumEmergent2015,nahumDeconfined2015,wangTensor-product2016,shaoQuantum2016,qinDuality2017,emidioNew2017,maDynamical2018,maRole2019,zhaoMulticritical2020,emidioDiagnosing2021,wangScaling2021,shuNonequilibrium2022,zhaoScaling2022} and have been gradually extended to interacting fermionic systems~\cite{satoDirac2017,liFermion-induced2017,liDeconfined2019a,liuMetallic2022,liuSuperconductivity2019,wangPhases2021,wangDoping2021}.
It is obvious that model design and large-scale quantum Monte Carlo (QMC) simulations played a key role in pushing the frontier of our knowledge on such surprising phenomena.

Except the DQCP discussed above,
interaction effects on massless Dirac fermions have also attracted great attentions.
Since the linear dispersion is stable against weak interactions, there must be one or more quantum phase transitions separating the Dirac semimetal (SM) phase and various possible Mott insulator states.
Depending on the type of interactions, many Mott insulators have been discovered, including the ferromagnetic and AF states~\cite{mengQuantum2010,changQuantum2012,otsukaUniversal2016,toldinFermionic2015,langQuantum2019,xuCompeting2021,moitabaChiral2021}, VBS state~\cite{assaadPhase2005,langDimerized2013,zhouMott2016,ouyangProjection2021,xuKekule2018,liaoValence2019}, nematic phase~\cite{schwabNematic2022}, superconductor and quantum (spin) Hall states~\cite{xuTopological2017,heDynamical2018,liuSuperconductivity2019,wangPhases2021,wangDoping2021} and many others.
Among these examples, particular interests lie in the direction where from Dirac SM to the strong-coupling limit, there exist multiple insulating phases as a function of the interaction strength, and leaves room for interesting intermediate phases such as topological ordered phases and multiple exotic quantum phase transitions, such as Gross-Neveu and DQCP.
Previous works have shown, with spin-1/2 electron and SU(2) symmetry, extended interaction on honeycomb lattice offer a robust VBS with Kekul\'e pattern exactly as such an intermediate phase between Dirac SM and strong-coupling AF order~\cite{xuKekule2018}. However, although the transition between Dirac SM with Kekul\'e VBS is found to be a Gross-Neveu QCP, the transition between VBS and AF phases is first order.
These results motivate us to investigate the interaction effect in $\pi$-flux extended Hubbard model on square lattice, as we show below, in this case, except a Gross-Neveu QCP between Dirac SM and VBS, there indeed further emerges a continuous transition between the VBS and AF phase within the largest system sizes accessed, consistent with the expected behavior of the DQCP. Our results of the Dirac fermion with an extended interaction, could also shed light on the great on-going efforts in understanding the interaction effects on quantum Moir\'e material models~\cite{liaoValence2019,liaoCorrelated2021,liaoCorrelation2021,zhangMomentum2021,panDynamical2022,zhangSuperconductivity2021,chenRealization2021,linExciton2022} such as twisted bilayer graphene (TBG) and transition metal dichalcogenides, where the interplay between flat band Dirac cones and the extended Coulomb interactions can be engineered by twisting angles, gating and tailored design of the dielectric environment, and give rise to a plethora of exotic phenomena.

{\it Model and Method.---}
We study a SU(2) extended Hubbard model with $\pi$-flux hopping term at half-filling on square lattice
 \begin{equation}
 \label{eq:model}
H=-\sum_{\langle i j\rangle, \sigma} t_{i j}\left(c_{i \sigma}^{\dagger} c_{j \sigma}+\text { H.c. }\right)+U \sum_{\square}\left(n_{\square}-1\right)^{2},
\end{equation}
where $\langle i j\rangle$ denotes the nearest neighbor,
$c_{i \sigma}^{\dagger}$ and $c_{i \sigma}$ are creation and annihilation operators for fermions on site $i$ with spin $\sigma=\uparrow, \downarrow$,
$n_\square$ is the extended particle number operator of $\square$-plaquette defined as $n_{\square}  \equiv \frac{1}{4} \sum_{i \in \square} n_{i} $ with $n_{i}= \sum_{\sigma} c_{i \sigma}^{\dagger} c_{i \sigma}$
and at half-filling $\langle n_{\square} \rangle =1$,
$U$ tunes the interaction strength, which favors AF order in the strong coupling limit~\cite{ouyangProjection2021}.

\begin{figure}[htp!]
\includegraphics[width=1\columnwidth]{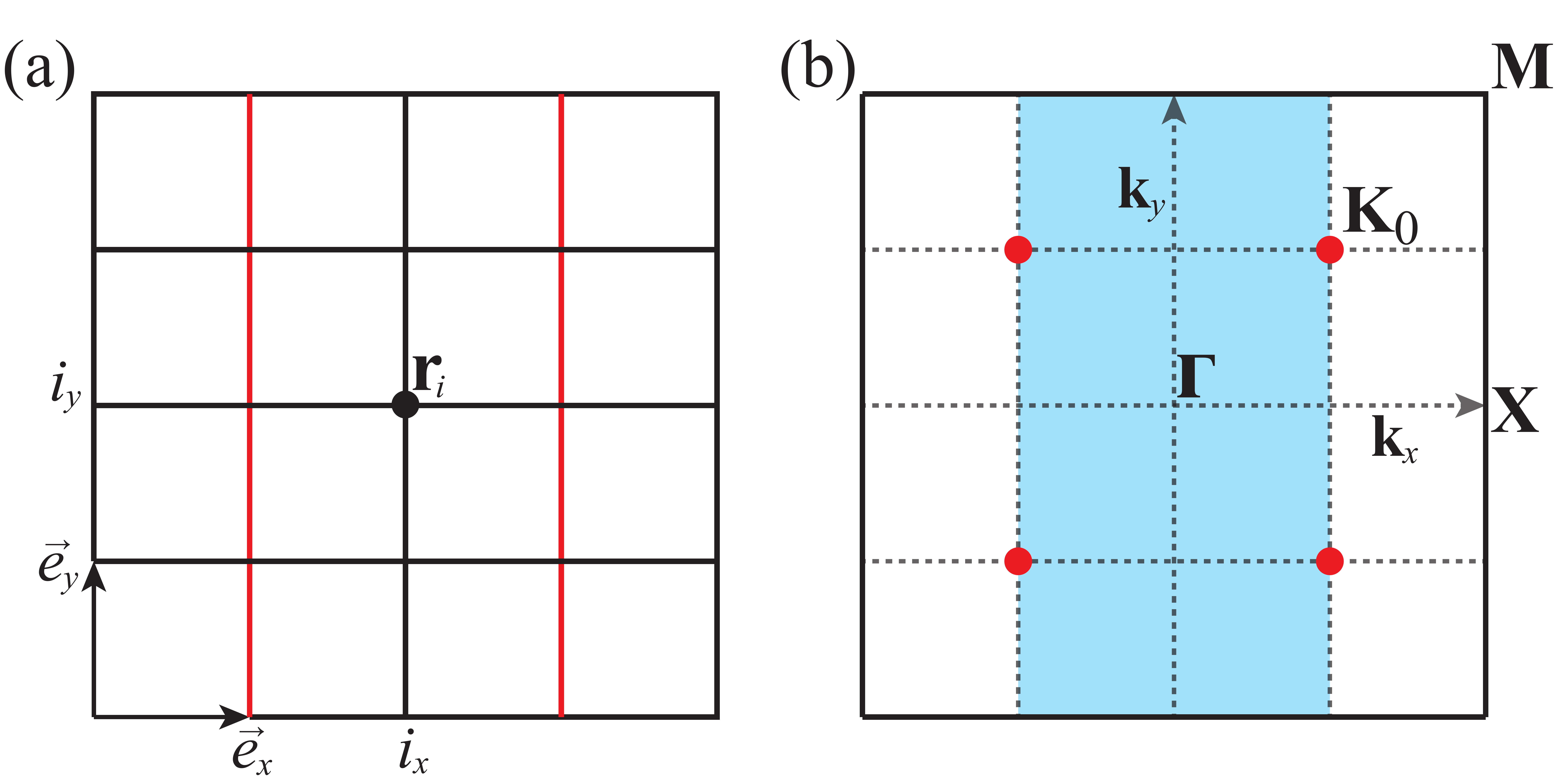}
\caption{(a) Square lattice with $\pi$-flux hopping. Red and black solid lines correspond to $-t$ and $t$. $\vec{e}_x$ ($\vec{e}_y$) is the unit vector along x (y) direction. The position of site $i$ is given as $\mathbf{r}_i = i_{x}\vec{e}_{x}+i_{y}\vec{e}_{y}$. (b) The white square is the BZ of the original square lattice, the blue BZ is the folded one considering the translation-symmetry-breaking hopping amplitudes. The red solid points represent the position of Dirac cones $\mathbf{K}_0 = \left( \frac{\pi}{2},\frac{\pi}{2}\right)$. High symmetry points $\mathbf{\Gamma}=(0,0)$, $\mathbf{X}=(\pi,0)$ and $\mathbf{M}=(\pi,\pi)$ are denoted.}
\label{fig:model}
\end{figure}

As shown in Fig.~\ref{fig:model}~(a), we set hopping amplitudes $t_{i,i+\vec{e}_{x}} = t$  and  $t_{i,i+\vec{e}_{y}} = (-1)^{i_x} t$, where the position of site $i$ is given as $\mathbf{r}_i = i_{x}\vec{e}_{x}+i_{y}\vec{e}_{y}$ and $t=1$ is the energy unit.
Such arrangement bestows a $\pi$-flux penetrating each $\square$-plaqutte.
As a consequence, two Dirac cones are located at $\mathbf{K}_0 = \left( \frac{\pi}{2}, \pm \frac{\pi}{2}\right)$ in the first Brillouin zone (BZ).
We note the way of the folding and locations of the Dirac cones all depend on the gauge choice of hopping amplitudes, i.e., with the above hopping the BZ is folded in half (the blue area in Fig.~\ref{fig:model}~(b)), but the distance between two Dirac cones is actually gauge invariant. The model therefore still has full crystalline symmetries of the square lattice (the p4mm wallpaper group),
where each crystalline symmetry operation is supplemented by a U(1) gauge transformation.
For example, the translation symmetry becomes $\hat{T}_{\vec{e}_x} : c_i \rightarrow (-1)^{i_y} c_{i+\vec{e}_x}$.
Consequently, we will still discuss our results in the original square lattice BZ.

For the extended Hubbard interaction term, the onsite, first and second nearest neighbor repulsions are all included in one plaquette.
This particular extended Coulomb interaction form can be related with %some kind of unknown
quantum Moir\'e materials with square lattice structure.
Because Wannier orbitals of Moir\'e materials, such as TBG, are quite extended, the relatively long-range Coulomb interactions have to be included to construct the effective model~\cite{koshinoMaximally2018,poOrigin2018}.
As found in the previous studies~\cite{xuKekule2018,liaoValence2019,liaoCorrelation2021,liaoCorrelated2021}, such extended interaction will require a relative larger $U$ to gap out the Dirac cones.

One can easily show the Hamiltonian in Eq.~\eqref{eq:model} is sign-problem free for auxiliary field QMC~\cite{wuSufficient2005} and we implement a projector version of QMC method~\cite{assaadWorld-line2008} to solve the model.
Details of the algorithm can be found in Supplemental Material (SM)~\cite{suppl}, and we only mention the projection length $\beta t = L$ for equal-time measurements; $\beta t  = L+10$ for imaginary-time measurements and discrete time slice $\Delta\tau =0.1$. We simulate the systems with linear size $L=12,16,20,24,28,32$.
We have also tested that this setup is enough to achieve controlled errorbars~\cite{suppl}.

\begin{figure}[htp!]
\includegraphics[width=1\columnwidth]{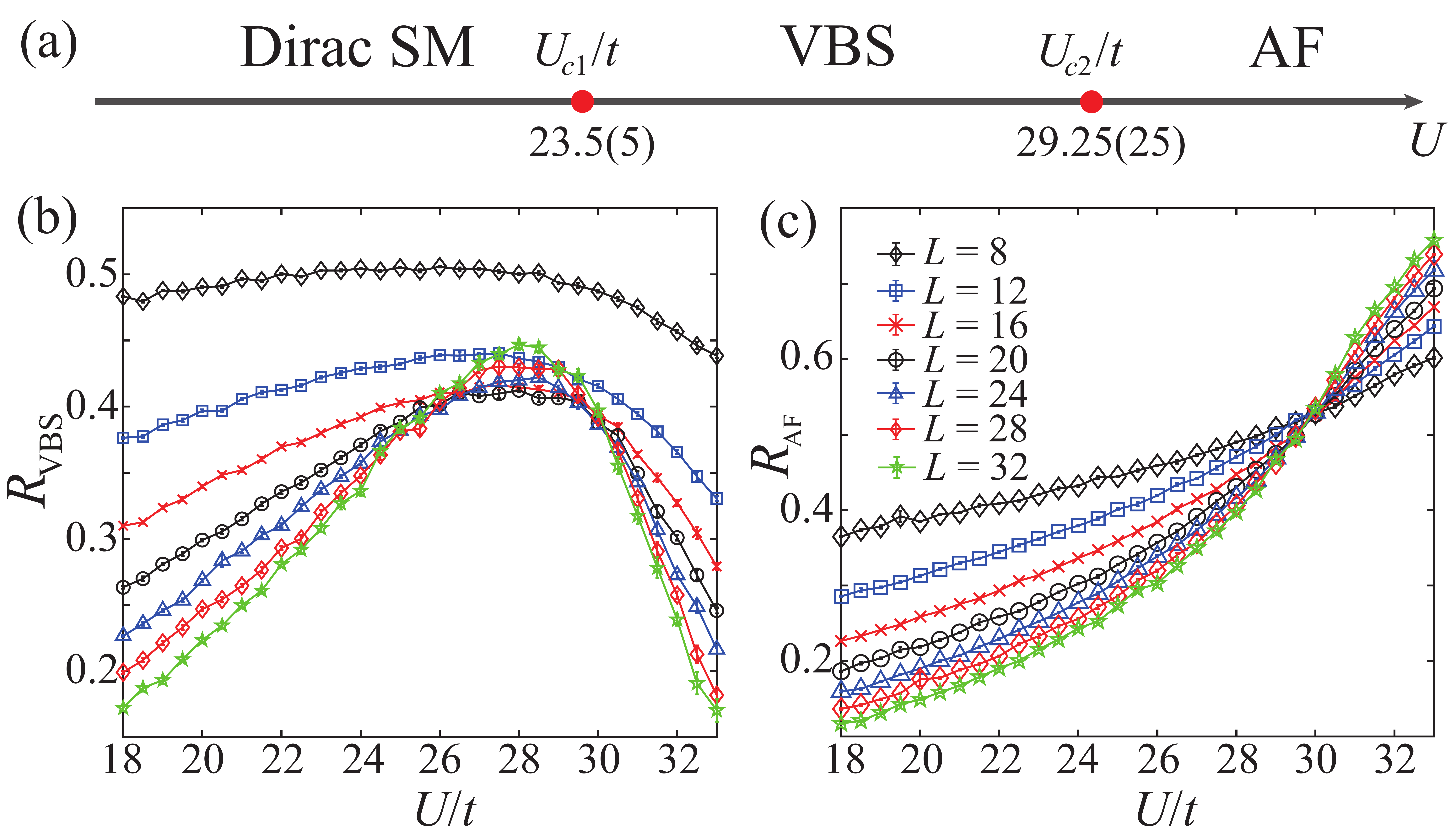}
\caption{(a) Phase diagram of the extended Hubbard model as function of interaction strength $U$, obtained from QMC simulations. The SM-VBS transition at $U_{c1}$ is continuous and of Gross-Neveu universality.
The VBS-AF transition at $U_{c2}$ is also continuous and should be explained according to the deconfined quantum criticality.
Correlation ratios of the (b) VBS order and (c) AF order as a function of interaction strength $U$ are shown.}
\label{fig:phasediagram}
\end{figure}

{\it QMC results }\,---\,
The phase diagram obtained from QMC simulations is shown in Fig.~\ref{fig:phasediagram}~(a).
%We find two quantum critical points when gradually increasing the interaction strength $U$.
We find an emergent intermediate phase which realizes two continuous quantum phase transitions when gradually increasing the interaction strength $U$.
They are the phase transition from Dirac SM to VBS phase and that from VBS to AF phase.
This particular sequence of transitions has not been observed in other models.
The first corresponding QCP is at $U_{c1}/t = 23.5(5)$ and of Gross-Neveu type with the VBS acquiring a $Z_{4}$ discrete lattice symmetry breaking, and the critical point is expected to have emergent U(1) symmetry as the $Z_4$ anisotropy is irrelevant~\cite{louAntiferromagnetic2009}. The second corresponding QCP is at $U_{c2}/t = 29.25(25)$, separating two spontaneous symmetry breaking phases, e.g. $Z_4$ for VBS and SU(2) for AF phases, and shall be explained according to the deconfined quantum criticality~\cite{senthilDeconfined2004,senthilQuantum2004}.
What's more, the corresponding critical exponents $\eta_\text{VBS}=0.6(1)$ and $\eta_\text{AF}=0.58(3)$ are extracted, they satisfy the CFT Bootstrap bound~\cite{nakayamaNecessary2016,poland2019conformal}.
%We note the existence of a DQCP in the Hubbard model on the square lattice has not been reported before.
In particular, to our best knowledge, our model is the first one-tuning-parameter fermionic model %with extended interaction
that gives rise to the critical exponents $\eta_\text{VBS}\approx \eta_\text{AF}$ meeting the CFT Bootstrap bound at DQCP .
%the Hubbard model with only onsite interactions hosts only Dirac SM and AF phases~\cite{otsukaUniversal2016,toldinFermionic2015}, and on the honeycomb lattice with extended interaction, the corresponding transition is first order~\cite{xuKekule2018}.

To quantitatively study the two phase transitions, we define two structure factors
\begin{equation}
C_\text{AF}(\mathbf{k}, L) = \frac{1}{L^4} \sum_{i,j} e^{i \mathbf{k} \cdot\left(\mathbf{r}_{i}-\mathbf{r}_{j}\right)}\left\langle \mathbf{S}_{i} \mathbf{S}_{j} \right\rangle
\label{eq:eq2}
\end{equation}
and
\begin{equation}
C_\text{VBS}(\mathbf{k}, L) = \frac{1}{L^4} \sum_{i,j} e^{i \mathbf{k} \cdot\left(\mathbf{r}_{i}-\mathbf{r}_{j}\right)} \left\langle B _{i} B _{j}\right\rangle
\label{eq:eq3}
\end{equation}
for AF order and VBS order, respectively.
In above equations,  $\mathbf{S}_{i}= \frac{1}{2} \sum_{\sigma \sigma^{\prime}} c_{i \sigma}^{\dagger} \boldsymbol{\sigma}_{\sigma \sigma^{\prime}} c_{i \sigma^{\prime}}$ are the fermion spin operators at site $i$, $\boldsymbol{\sigma}$ the Pauli matrices.
$ B_{i} = \sum_{\sigma}\left( t_{i, i+\vec{e}_x}  c_{i, \sigma}^{\dagger} c_{i+\vec{e}_x, \sigma}+ \text {H.c.} \right)$ are gauge invariant bond operators.
For AF order, $C_\text{AF}(\mathbf{k}, L)$ is peaked at momentum $\mathbf{M}=(\pi,\pi)$;
for VBS order, $C_\text{VBS}(\mathbf{k}, L)$ is peaked at momentum $\mathbf{X}=(\pi,0)$.
We then use the renormalization-group invariant correlation ratios ($c = \text{VBS}, \text{AF}$) to perform the data analysis,
\begin{equation}
	R_{c}(U, L)=1-\frac{C_{c}\left(\mathbf{k}=\mathbf{Q}+d\mathbf{q},L\right)}{C_{c}\left(\mathbf{k}=\mathbf{Q},L\right)},
\end{equation}
where $\mathbf{Q}$ is the ordering wave vector, $|d\mathbf{q}| \sim \frac{1}{L}$ denotes the smallest momentum on finite size lattice.
By definition, $R_{c}(U, L)\rightarrow 1$ ($0$) for $L\rightarrow \infty$ in the corresponding ordered (disordered) phase.
At the QCP, $R_{c}$ is scale invariant for sufficiently large $L$ and exhibit the scaling behavior for the corresponding universalities~\cite{campostriniFinite-size2014,kaulSpin2015,pujariInteraction2016,langQuantum2019,liDeconfined2019a}.

As shown in Fig.~\ref{fig:phasediagram}~(b), when varying $U/t$ from $18$ to $33$, $R_\text{VBS}(U,L)$ first increases then decreases. Importantly, $R_\text{VBS}(U,L=20,24,28,32)$ for different $L$ cross at two separate regions.
These results mean that our model undergoes two phase transitions, and the VBS order is the intermediate phase. Admittedly, the VBS order is very weak but remain finite at thermodynamic limit (TDL), and we believe it is attributed to the enhanced quantum fluctuations from the interplay of Dirac fermions and extended interactions within a plaquette. We also find at the transition, the VBS order parameter histogram is consistent with emergent U(1) symmetry~\cite{xuKekule2018}.
The correlation ration of AF order is shown in  Fig.~\ref{fig:phasediagram}~(c).
Interestingly, a clear crossing takes place around $U/t=29.4$, which further indicates the phase transition between the VBS and the AF order is continuous.

\begin{figure}[htp!]
\includegraphics[width=1\columnwidth]{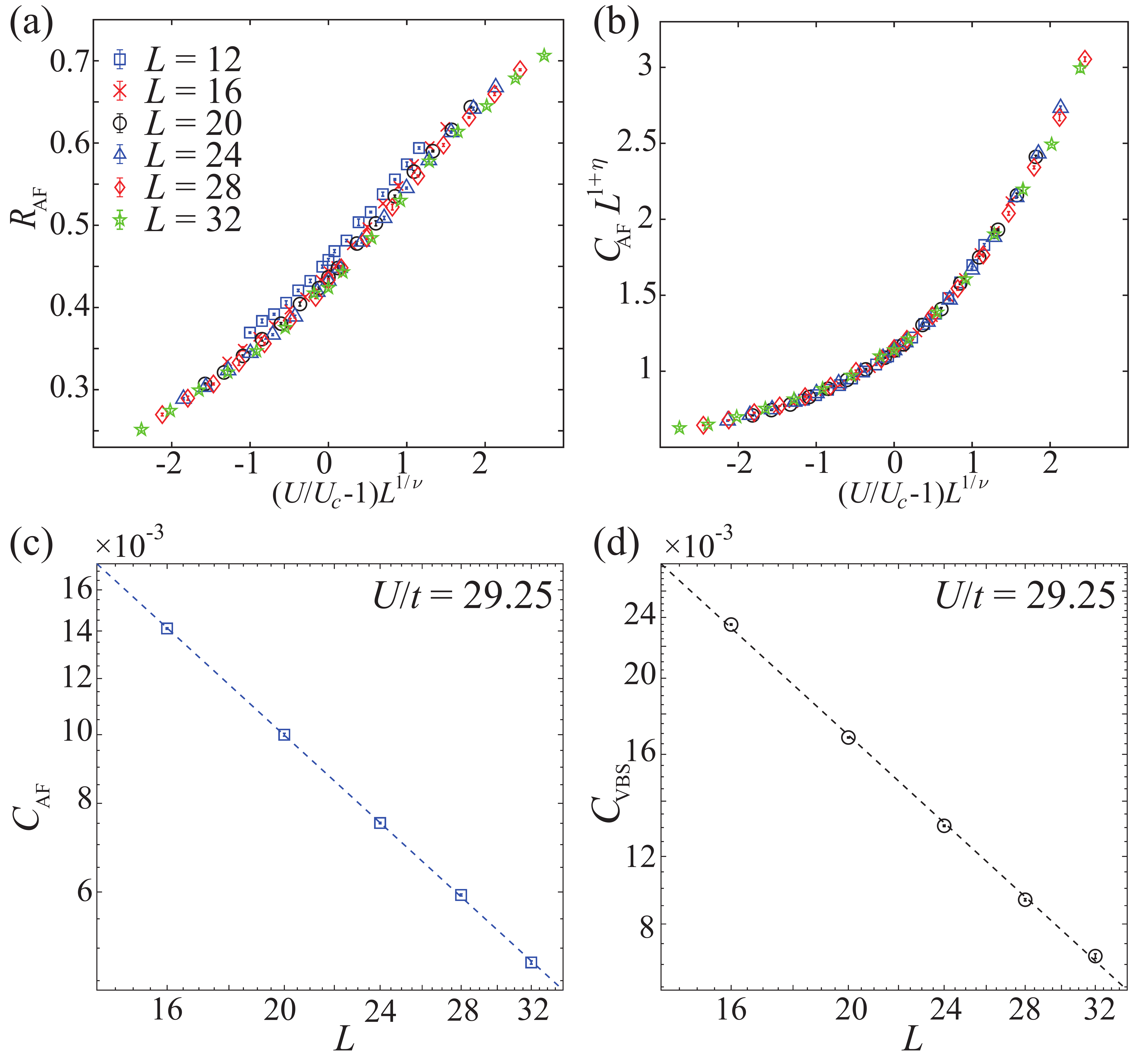}
\caption{(a) The data collapse of correlation ratio $R_\text{AF}$, which gives $U_{c2}/t = 29.25(25)$ and $\nu=1.13(5)$.
(b) The data collapse of AF structure in the vicinity of $U_{c2}/t=29.25$ with $\nu=1.13$ and $\eta = 0.58$.
The log-log plot of the structure factor of (c) AF order and (d) VBS order versus the linear lattice size $L$ at $U_{c2}/t=29.25$. The critical exponents $1+\eta$ can be extracted from the slopes of linear fitting curves in log-log plots. We obtain $\eta_\text{AF} = 0.58(3)$ and $\eta_\text{VBS}=0.6(1)$.}
\label{fig:VBS-AF}
\end{figure}

To understand the two intriguing QCPs, we first focus on the more complicated VBS-AF transition, to demonstrate it has the flavor of DQCP. To this end, we collapse the correlation ratio of AF order with finite-size scaling relation $R_\text{AF}(U, L)=f_{1}\left(L^{1 / \nu} \left(U/ U_{c}-1\right) \right)$, as shown in Fig.~\ref{fig:VBS-AF}~(a), and obtain the position of corresponding QCP at $U_{c2}/t=29.25(25)$ and the correlation length exponent $\nu=1.13(5)$.
Then at $U_{c2}$ the AF and VBS structure factors, in Eqs.~\eqref{eq:eq2} and \eqref{eq:eq3}, shall obey the following finite-size scaling ansatz~\cite{liaoValence2019,zhouMott2018}
\begin{equation}\label{eq:structure-factor-collapse-ansatz}
C_{c}(U, L)=L^{-z-\eta}f_{2}\left(L^{1 / \nu} \left(U-U_{c}\right) / U_{c} \right),
\end{equation}
where $\eta$ is the anomalous dimension exponent, $z=1$ is the dynamic exponent.
As shown in Fig.~\ref{fig:VBS-AF}~(c) and (d), we extract $\eta$ from the slope of log-log plot of $C_{c}(U, L)$ curves at $U_{c2}/t=29.25$ and find $\eta_\text{AF} =0.58(3)$ for AF order and $\eta_\text{VBS} =0.6(1)$ for VBS order, respectively, give rise to good quality linear fits.

According to the scenario of DQCP~\cite{senthilQuantum2004,senthilDeconfined2004}, the closeness of the exponents from the two ordered phase approaching the critical point, i.e. $\eta_\text{VBS} \approx \eta_\text{AF}$ in our setting, is considered as an important signature for the associated emergent symmetry~\cite{nahumEmergent2015,nahumDeconfined2015}, and numerical evidences of such closenesses have been seen in the J-Q model~\cite{sandvikEvidence2007,qinDuality2017}, 3D loop model~\cite{nahumDeconfined2015}, as well as those in the recently discovered the fermionic models~\cite{liuSuperconductivity2019,liDeconfined2019a,wangPhases2021,wangDoping2021}.
In the literature~\cite{nahumEmergent2015,nahumDeconfined2015,qinDuality2017,liuSuperconductivity2019,liDeconfined2019a,wangPhases2021,wangDoping2021}, the precise value of the exponents appears to depend on the detailed implementation of each model, and there also exists the conformal field theory (CFT) bound that the emergent symmetry need to satisfy~\cite{nakayamaNecessary2016,poland2019conformal}.       We note the $\eta_\text{VBS}$ and $\eta_\text{AF}$ in our work satisfy the CFT Bootstrap bound, as well as in a completely different interacing fermion model on a honeycomb lattice in Ref.~\cite{liDeconfined2019a}. Importantly, comparing with Ref.~\cite{liDeconfined2019a}, we only use one tuning parameter in our model, as there is no interference from nearby multicriticality or first-order transition as in Ref.~\cite{liDeconfined2019a}, which may pollute the measurement of critical exponents because a clean scaling behavior can only be observed far away from the multicritical point or first-order transitions.
The more recent entanglement measurements further point out the DQCP, at least in the J-Q model, might not be an unitary CFT at the first place~\cite{wangScaling2021,zhaoScaling2022} and other possible scenarios such as multicritical point~\cite{zhaoMulticritical2020}, complex CFT~\cite{maTheory2020,nahumNote2020} and weakly first order transition~\cite{emidioDiagnosing2021}, have been constantly and actively proposed and explored. Despite of these efforts and the enigmatic situation of the DQCP, our observation, that within the system sizes simulated, $\eta_\text{VBS} \approx \eta_\text{AF}$, is consistent with the deconfinement at $U_{c2}$. It is further worthwhile to point out that a relative large $\eta \sim 0.6$ is also the hallmark of many QCPs associated with the fractionaliazation of elementary excitations such as the condensation transition of spinon and visons in $Z_N$ topological orders~\cite{IsakovUniversal2012,wangQuantum2018,wangFractionalized2020}.
We also collapse the AF structure factor according to the Eq.~\eqref{eq:structure-factor-collapse-ansatz} with $U_{c2}/t=29.25$, $\nu=1.13$ and $\eta=0.58$, as shown in Fig.~\ref{fig:VBS-AF}~(b), all data points fall on a smooth curve.
Therefore, our numerical data in Fig.~\ref{fig:VBS-AF} certainly reveal an internally consistent decription along the the line of DQCP for the VBS-AF transition.

\begin{figure}[htp!]
\includegraphics[width=1\columnwidth]{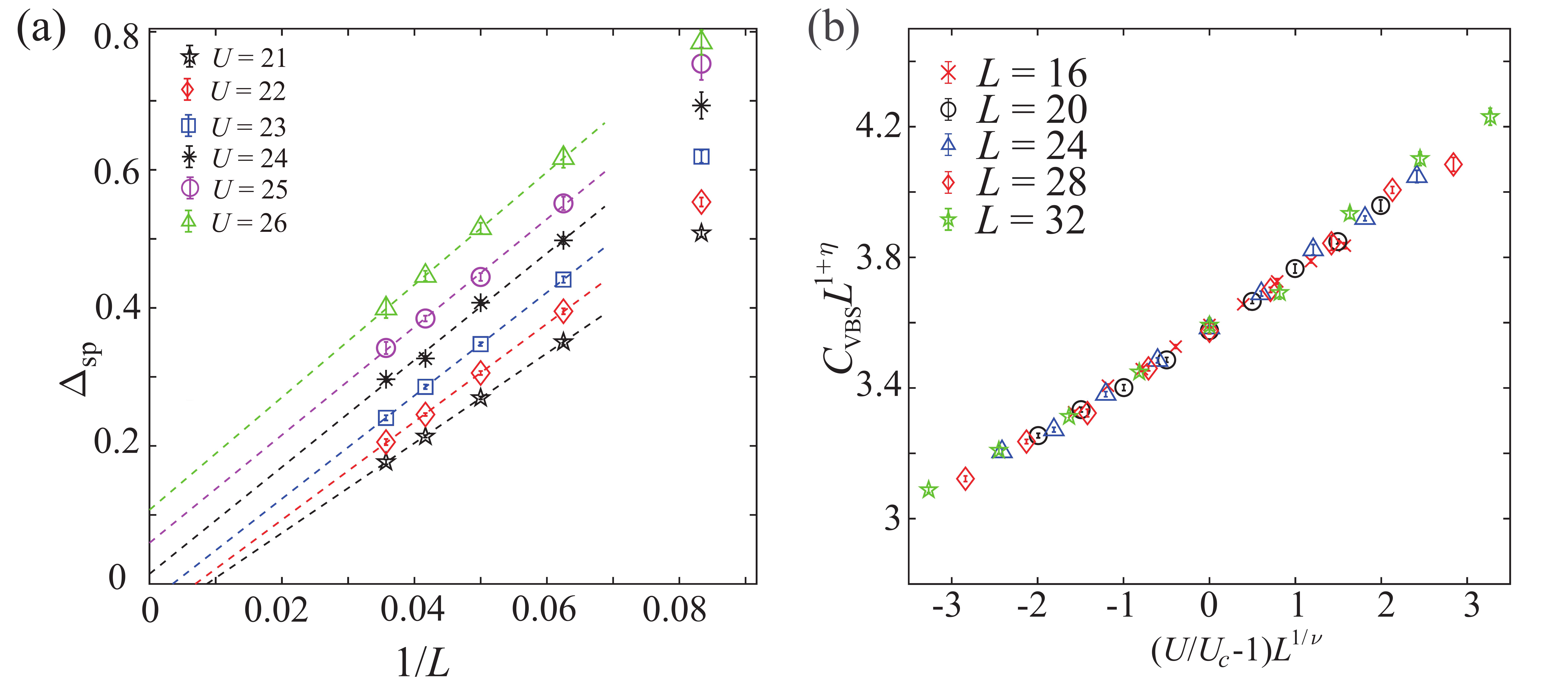}
\caption{(a) The $1/L$ extrapolation of the single-particle gap $\Delta_{\mathrm{sp}}(\mathbf{K}_0,L)$, the gap opens at an interaction strength locating in a range from $U/t=23$ to $U/t=24$.
(b) The data collapse of the structure factor of VBS order, which gives an estimation of $U_{c1}/t =23.5(5)$, $\nu^\prime =1.0(1)$ and $\eta^\prime=0.89(3)$.
}
\label{fig:SM-VBS}
\end{figure}

Next we move on to the SM-VBS transition. It is known that the massless Dirac fermion is robust at weak interaction, and the single-particle gap will open at a finite interaction strength~\cite{liuMetallic2022,liaoCorrelated2021}.
In our model, we indeed find as a function of $U$, the Dirac SM transit to an insulating VBS order continuously through a Gross-Neveu QCP
~\cite{gross1974,hands1993,rosenstein1993critical,Zerf2017,liFermion-induced2017,zhouMott2016,scherer2016gauge,mihaila2017gross,jian2017fermion,classen2017fluctuation,torres2018fermion,Ihrig2018,xuKekule2018,zhouMott2018,liaoValence2019}.
This is also consistent with the similar situation of the extended Hubbard model on honeycomb lattice~\cite{xuKekule2018,liaoValence2019}.

To determine $U_{c1}$, we extract the single-particle gap $\Delta_{\mathrm{sp}}(\mathbf{K}_0,L)$ from a fit to the asymptotic long imaginary time behavior of the single-particle Green's function
$G(\mathbf{k}, \tau, L)=\left(1/L^{4}\right) \sum_{i, j, \sigma} e^{i \mathbf{k} \cdot\left(\mathbf{r}_{i}-\mathbf{r}_{j}\right)}\left\langle c_{i, \sigma}^{\dagger}(\tau) c_{j, \sigma}(0)\right\rangle \propto e^{-\Delta_{\mathrm{sp}}(\mathbf{k},L) \tau}$.
The obtained $\Delta_{\mathrm{sp}}(\mathbf{K}_0,L)$ are shown in Fig.~\ref{fig:SM-VBS}~(a).
It is clear that $\Delta_{\mathrm{sp}}(\mathbf{K}_0,L\rightarrow\infty) \rightarrow 0$ at $U/t<23$ and $\Delta_{\mathrm{sp}}(\mathbf{K}_0,L\rightarrow\infty) > 0$ at $U/t>24$, which indicates $U_{c1}/t\in(23,24)$ and is overall consistent with the cross point of $R_\text{VBS}$ shown in Fig.~\ref{fig:phasediagram}~(b). This again signifies the weakness of the VBS order and the strong fluctuations at this QCP which give rise to strong finite-size effect.
To locate the $U_{c1}$ more accurately, as shown in Fig.~\ref{fig:SM-VBS}~(b), we collapse the VBS structure factor according to the Eq.~\eqref{eq:structure-factor-collapse-ansatz} in $U/t\in (23, 24)$.
Although the finite-size effect is strong, the data collapse nevertheless gives rise to an estimation, $U_{c1}/t=23.5(5)$, and critical exponents, $\nu^\prime=1.0(1)$ and $\eta^\prime=0.89(3)$.
These exponents are consistent with previous numerical simulations of similar SM-VBS transitions on the honeycomb lattice~\cite{xuKekule2018}, where it is found the three-fold lattice rotation symmetry is enhanced to an emergent U(1) at the Gross-Neveu QCP. Since the three-fold anisotropy of the U(1) order parameter is (dangerously) irrelevant at the QCP, it is expected that the four-fold anisotropy should be even more irrelevant in our case.

\begin{figure}[htp!]
\includegraphics[width=1\columnwidth]{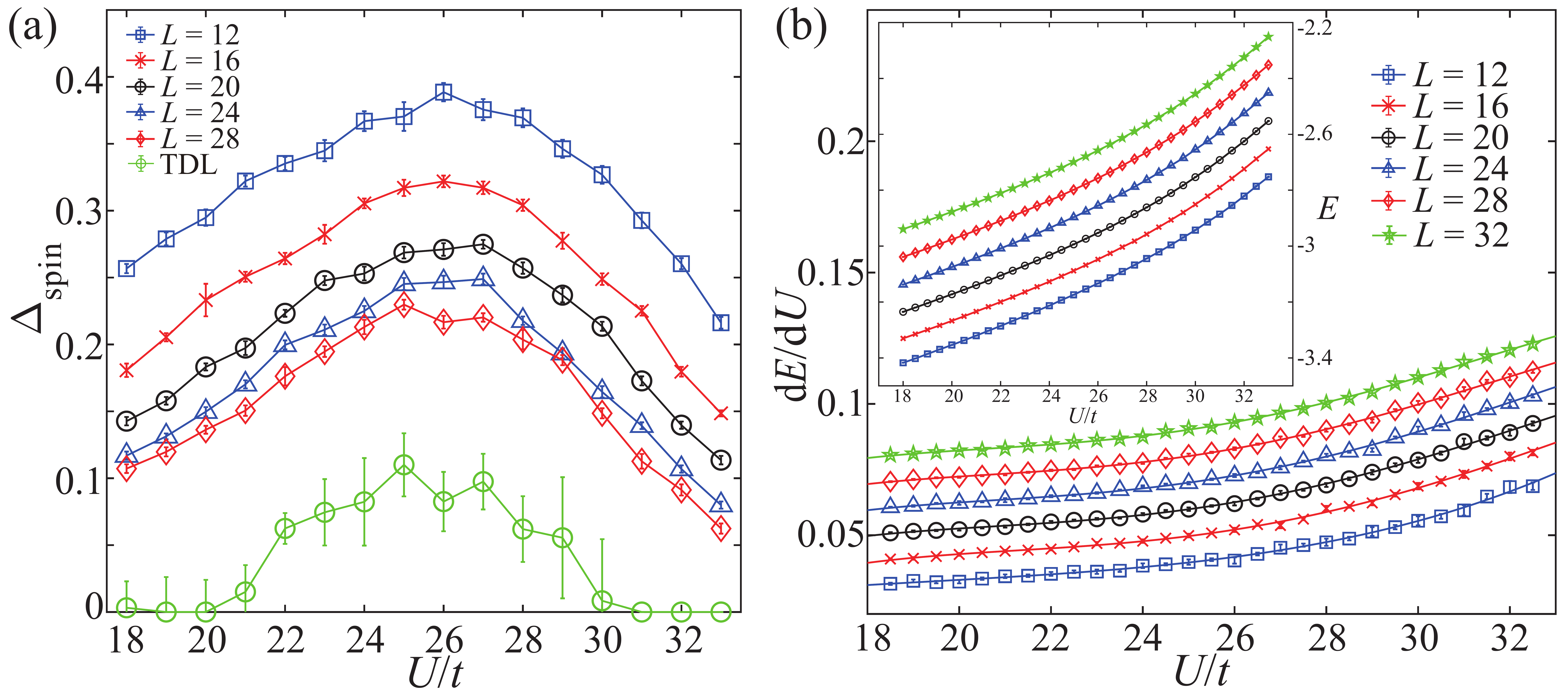}
\caption{(a) Spin excitation gap $\Delta_{\mathrm{sp}}(\mathbf{K}_0)$ for different $L$ and their TDL extrapolation as a function of interaction strength $U$. The VBS phase has a finite spin gap due to the formation of singlets.
(b) Main panel: first derivative of kinetic energy density as a function of $U$, the solid curves are a cubic polynomial fitting through the data. No discontinuity observed. Inset:  kinetic energy density as a function of $U$. Lines and points in both main panel and inset are shifted for visualization purpose without changing the physical meaning. Error bars are smaller than the symbols.}
\label{fig:spingap}
\end{figure}

At $U<U_{c1}$, the ground state is in Dirac SM state, and thus there is no spin excitation gap in the TDL.
At $U>U_{c2}$, the ground state is in AF state, the spin excitation gap in the TDL should also vanish because of the existence of Goldstone mode. However, the spin excitation gap will open in the VBS state due to the formation of a spin singlet~\cite{liaoValence2019}.
To verify the theoretical predictions, we measure the dynamical spin-spin correlation function
$C(\mathbf{k}, \tau, L)=\frac{1}{L^4} \sum_{i, j} e^{i \mathbf{k} \cdot\left(\mathbf{r}_{i}-\mathbf{r}_{j}\right)}\left\langle \mathbf{S}_{i}(\tau) \mathbf{S}_{j}(0)\right\rangle$.
The spin excitation gap $\Delta_{\mathrm{spin}}\left(\mathbf{M},L \right)$ can be extract from  imaginary-time decay of $C(\mathbf{k}, \tau, L) \propto e^{-\Delta_{\mathrm{spin}}\left(\mathbf{k},L\right)\tau}$.
As shown in Fig.~\ref{fig:spingap}~(a) we extrapolate spin excitation gap to the TDL and find that $\Delta_{\mathrm{spin}}\left(\mathbf{M}\right)$ goes to a finite value near $U_{c1}$ and goes back to zero near $U_{c2}$.
These QMC results are consisted with our above theoretical analysis, and thus confirm the process of evolution of the ground state of our model, \emph{i.e.} transiting from Dirac SM to VBS state first and then from VBS to AF state,  as a function of interaction strength $U$.

In addition, we provide more evidence that the two QCPs are continuous within the system size studied, by means of monitoring the evolution of the expectation value of the kinetic energy density. Since our QMC is of projector type, this is meant to monitor the evolution of the (part of) the free energy of the system. As shown in the inset of Fig.~\ref{fig:spingap}~(b), the kinetic energy density $E = \frac{1}{L^4}\left\langle-\sum_{\langle i j\rangle, \sigma} t_{i j}\left(c_{i \sigma}^{\dagger} c_{j \sigma}+\text { H.c. }\right)\right\rangle$ evolve smoothly as function of $U$ for different system sizes. We then compute their first order derivatives and find in the main panel of Fig.~\ref{fig:spingap} (b) no discontinuities, consistent with the continuous phase transition.
In the SM~\cite{suppl}, we also present similar analysis of the structure factors for AF and VBS orders. They support the SM-VBS and VBS-AF transition are all continuous.

{\it Discussions }\,---\,
With the help of large-scale sign-free projector QMC simulation, we investigate the phase diagram of $\pi$-flux extended Hubbard model on square lattice at half-filling.
Based on all the numerical results obtained, we conclude that this simple looking model accquire an interesting phase diagram with an intermediate phase with weak VBS order separating the well known Dirac SM and AF phases.
More importantly, we find the Gross-Neveu transition from Dirac SM to VBS is continuous, and the transition from VBS to AF is also continuous, consistent with deconfined quantum critical criticality.

Our results, along with the previous works of extended interaction model on honeycomb lattice~\cite{xuKekule2018,liaoValence2019,ouyangProjection2021}, point out the directions that to realize interesting phase diagrams with (multiple) intermediate phases between the free Dirac SM and strong coupling Mott insulators, the extended interactions beyond on-site Hubbard term are crucial and could bring more surprises. In fact, the weak VBS order discovered here and the DQCP associated with it towards the AF order, imply further perturbations could  give rise to even more exotic interaction-driven phases and transitions. In this context, our results also have the relevance to the great on-going efforts in understanding the interaction effect on quantum Moir\'e material models~\cite{xuKekule2018,liaoValence2019,ouyangProjection2021,liaoCorrelated2021,liaoCorrelation2021,zhangMomentum2021,panDynamical2022,zhangSuperconductivity2021} such as twisted bilayer graphene and transition metal dichalcogenides, where the interplay between flat band Dirac cones and the extended Coulomb interactions can be engineered by twisting angles, gating and tailored design of the dielectric environment, and give rise to a plethora of exotic phenomena and interesting phase and phase transitions. It is natural to anticipate, with the technique and analysis presented in this work, once further degrees of freedom and tunabilities in Moir\'e systems are added, interesting phases and transitions and their mechanism will be revealed from the lattice model simulations.

{\it Note Add}\,---\,
After the completion of this work, we become aware of a related investigation by Zhu et al~\cite{zhuQuantum2022}, in which consistent results are obtained.

{\it Acknowledgements }\,---\,
Y.D.L. acknowledges the support of Project funded by China Postdoctoral Science Foundation through Grants No. 2021M700857 and No. 2021TQ0076.
X.Y.X. is sponsored by the National Key R\&D Program of China (Grant No. 2021YFA1401400), Shanghai Pujiang Program under Grant No. 21PJ1407200, Yangyang Development Fund, and startup funds from SJTU. \
Z.Y.M. acknowledges support from the Research Grants Council of Hong Kong SAR of China (Grant Nos. 17303019, 17301420, 17301721 and AoE/P-701/20), the GD-NSF (No. 2022A1515011007), the K. C. Wong Education Foundation (Grant No. GJTD-2020-01) and the Seed Funding “Quantum-Inspired explainable-AI” at the HKUTCL Joint Research Centre for Artificial Intelligence.
Y.Q. acknowledges support from the the National Natural Science Foundation of China (Grant Nos. 11874115 and 12174068).
The authors also acknowledge \href{https://www.paratera.com/}{Beijng PARATERA Tech Co.,Ltd.} for providing HPC resources that have contributed to the research results reported within this paper.

\bibliographystyle{apsrev4-2}
\bibliography{main}

\clearpage
\onecolumngrid

\begin{center}
\textbf{\large Supplemental Material for "Dirac fermions with plaquette interactions. I. SU(2) phase diagram with Gross-Neveu and deconfined quantum criticalities"}
\end{center}
%\appendix
\setcounter{equation}{0}
\setcounter{figure}{0}
\setcounter{table}{0}
\setcounter{page}{1}
\makeatletter
\renewcommand{\theequation}{S\arabic{equation}}
\renewcommand{\thefigure}{S\arabic{figure}}
\renewcommand{\bibnumfmt}[1]{[S#1]}
\renewcommand{\citenumfont}[1]{S#1}
\setcounter{secnumdepth}{3}

\section{Projection QMC method and Simulation details}
Since we are interested in the ground state properties, the projection quantum Monte Carlo (PQMC) method is a suitable choice. 
In PQMC, we calculate the ground state wave function $\vert \Psi_0 \rangle$ through the projection of a trial wave function $\vert \Psi_T \rangle$ {\it i.e.} $\vert \Psi_0 \rangle = \lim\limits_{\beta \to \infty} e^{-\frac{\beta}{2} \mathbf{H}} \vert \Psi_T \rangle$~\cite{assaadWorld-line2008}. 
Physical observables can be measured according to
\begin{equation}
\label{eq:observablepqmc}
\langle \hat{O} \rangle = \frac{\langle \Psi_0 \vert \hat{O} \vert \Psi_0 \rangle}{\langle \Psi_0 \vert \Psi_0 \rangle} 
						= \lim\limits_{\beta \to \infty} \frac{\langle \Psi_T \vert  e^{-\frac{\beta}{2} \mathbf{H}} \hat{O}  e^{-\frac{\beta}{2} \mathbf{H}} \vert \Psi_T \rangle}{\langle \Psi_T \vert  e^{-\beta \mathbf{H}} \vert \Psi_T \rangle} .
\end{equation}
where $\hat{O}$ is the concerned physical observable, $\mathbf{H}$ is the Hamiltonian operator, $\beta$ is the projection length.
Since $\mathbf{H}$ consists of the non-interacting and interacting parts, $K$ and $V$, respectively, that are usually not commute, we should performed Trotter decomposition to discretize $\beta$ into $L_\tau$ slices ($\beta=L_\tau \Delta\tau$), then
\begin{equation}
\langle\Psi_{T}|e^{-\beta H}|\Psi_{T}\rangle=\langle\Psi_{T}|\left(e^{-\Delta\tau H_{U}}e^{-\Delta\tau H_{0}}\right)^{L_\tau}|\Psi_{T}\rangle+\mathcal{O}(\Delta{\tau}^{2})
\end{equation}
where the non-interacting and interacting parts of the Hamiltonian is separated. 
It is worth to note that above processing will give rise to a small systematic error $\mathcal{O}(\Delta\tau^2)$, thus we should set $\Delta\tau$ as a small number.
In Hubbard model, the interaction parts contain the quartic fermionic operator that can not be measured directly with PQMC method. 
To calculate the interacting part, one need to employ a SU(2) symmetric Hubbard-Stratonovich (HS) decomposition, then the auxiliary fields will couple to the charge density.
For example, in our extended Hubbard model on square lattice at hall-filing, the HS decomposition is
\begin{equation}
e^{-\Delta\tau U(n_{\square}-1)^{2}}=\frac{1}{4}\sum_{\{s_{\square}\}}\gamma(s_{\square})e^{\alpha\eta(s_{\square})\left(n_{\square}-1\right)}
\label{eq:decompo}
\end{equation}
with $\alpha=\sqrt{-\Delta\tau U}$, $\gamma(\pm1)=1+\sqrt{6}/3$,
$\gamma(\pm2)=1-\sqrt{6}/3$, $\eta(\pm1)=\pm\sqrt{2(3-\sqrt{6})}$,
$\eta(\pm2)=\pm\sqrt{2(3+\sqrt{6})}$ and the sum symbol is taken over the auxiliary fields $s_{\square}$ on each square plaquette.

In the end, we derive out the following formula with constant factors omitted
\begin{widetext}
   \begin{eqnarray}
\langle\Psi_{T}|e^{-\beta H}|\Psi_{T}\rangle=\sum_{\{s_{\square,\tau}\}}\left[\left(\prod_{\tau}\prod_{\square}\gamma(s_{\square,\tau})e^{-\alpha\eta(s_{\square,\tau})}\right)\det\left[P^{\dagger}B(\beta,0)P\right]\right]
\label{eq:mcweight}
   \end{eqnarray}
\end{widetext}
here $P$ is the coefficient matrix of trial wave function $|\Psi_T\rangle$; $B$ matrix is defined as
$B(\tau+1,\tau)=e^{V_\tau}e^{-\Delta_\tau K}$
and has a property $B(\tau_3,\tau_1)=B(\tau_3,\tau_2)B(\tau_2,\tau_1)$.
In the practice,  we choose the ground state wavefunction of the half-filled non-interacting parts of Hamiltonian as the trial wave function. 
The Monte Carlo sampling of auxiliary fields are further performed based on the weight defined in the sum of  Eq.~\eqref{eq:mcweight}. The measurements are performed near $\tau=\beta/2$. Single particle observables are measured by Green's function directly and many body correlation functions are measured from the products of single-particle Green's function based on their corresponding form after Wick-decomposition. 
The equal time Green's function are calculated as
\begin{equation}
G(\tau,\tau)=1-R(\tau)\left(L(\tau)R(\tau)\right)^{-1}L(\tau)
\end{equation}
with $R(\tau)=B(\tau,0)P$, $L(\tau)=P^{\dagger}B(\beta,\tau)$. 
The imaginary-time displaced Green's function  
$G(\tau,0) \equiv \left\langle \mathbf{c} \left(\frac{\tau}{2}\right) \mathbf{c}^{\dagger}\left(-\frac{\tau}{2}\right)\right\rangle$
are calculated as
\begin{widetext}
   \begin{eqnarray}
%\begin{equation}
\left\langle \mathbf{c} \left(\frac{\tau}{2}\right) \mathbf{c}^{\dagger}\left(-\frac{\tau}{2}\right)\right\rangle=\frac{\left\langle\Psi_{T}\left|e^{-\left(\frac{\beta}{2}+\frac{\tau}{2}\right) H} \mathbf{c} e^{-\tau H} \mathbf{c}^{\dagger} e^{-\left(\frac{\beta}{2}-\frac{\tau}{2}\right) H}\right| \Psi_{T}\right\rangle}{\left\langle\Psi_{T}\left|e^{-\beta H}\right| \Psi_{T}\right\rangle}
%\end{equation}
   \end{eqnarray}
\end{widetext}
where $\mathbf{c} = \{c_1, c_2 \cdots c_{N} \}$, $N=L\times L$ is the system size.
More technique details of PQMC method, please refer to Refs~\cite{assaadWorld-line2008}. 

\begin{figure}[htp!]
\includegraphics[width=1\columnwidth]{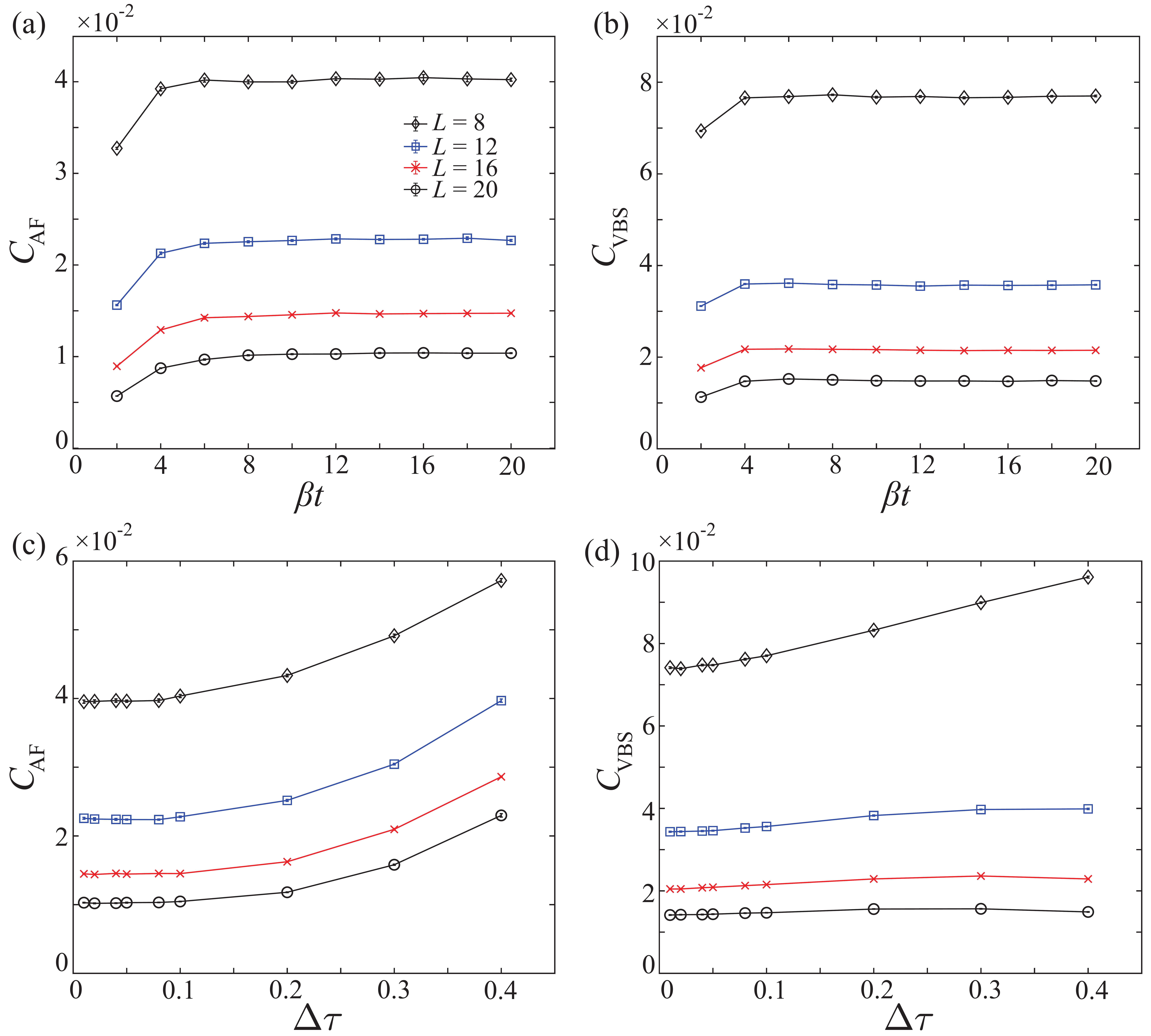}
\caption{The structure factors of (a) AF order and (b) VBS order with respect to the projection length $\beta$ with $\Delta\tau=0.1$. 
The structure factors of (c) AF order and (d) VBS order with respect to the time slice $\Delta\tau$ with $\beta t =L$. These results are obtained at $U/t=29.5$ near $U_{c2}$. All lines in the figure are guides to the eye.
}
\label{fig:QMCsetup}
\end{figure}

As mentioned in our paper, we set projection time $\beta t=L$ for equal-time measurement, $\beta t = L+10$ for imaginary-time measurement and discrete time slice $\Delta\tau=0.1$.
As shown in Fig~\ref{fig:QMCsetup}, we have confirmed that this setup is enough to achieve convergent and error controllable numerical results for our model.

\section{The smooth of structure factors}
As shown in previous QMC studies~\cite{xuKekule2018,liaoValence2019}, physical observable, like kinetic energy, structure factors, correlation ratios and so on, will present different behavior depending on the different phase transition types, i.e. continuous or first-order.
For the first-order phase transition, the curve of physical observable verse interaction strength will show a kink or jump, while for continuous one, smooth.
We have shown the smooth behavior of kinetic energy as varying interaction strength in main text.
Here, as shown in the Fig~\ref{fig:smoothcurves}~(a) and (b), we notice that the curves of structure factors are also smooth, which support the SM-VBS and VBS-AF transition all are continuous.

\section{The VBS order is weak but robust}
\begin{figure}[htp!]
\includegraphics[width=1\columnwidth]{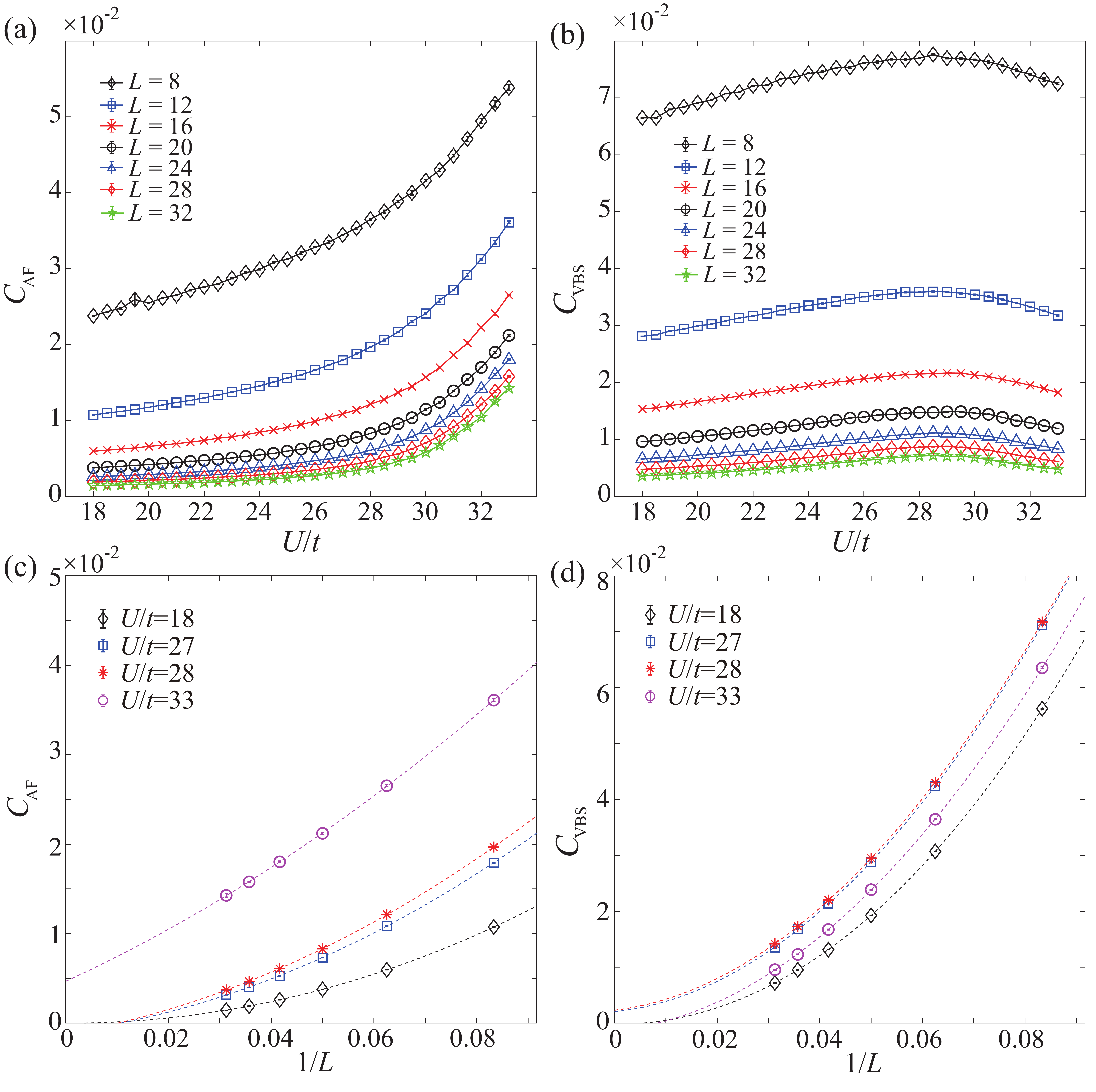}
\caption{The structure factors of (a) AF order and (b) VBS order as function of interaction strength $U$. All lines in the figure are guides to the eye, we can notice they show no jump, kink or discontinuous.
$1/L$ extrapolation of (c) $C_\text{AF}$ and (d) $C_\text{VBS}$.
}
\label{fig:smoothcurves}
\end{figure}

What's more, we extrapolate $C_\text{AF}$ and $C_\text{VBS}$ to thermodynamic limit at different $U$, as shown in Fig.~\ref{fig:smoothcurves}~(c) and (d).
We find that $C_\text{VBS}(1/L\rightarrow 0)$ at $U/t=27$ and $28$ are small but non-zero values, which means that the VBS order is weak but robust in our model. 
The $1/L$ extrapolation results are consistent with the positions of QCP and DQCP we extracted in main text.

\end{document}